\documentclass[aps,prd,onecolumn,amsmath,showpacs,superscriptaddress,nofootinbib,nopreprintnumbers,tightenlines,notitlepage]{revtex4-1}
\usepackage{verbatim}
\usepackage[T1]{fontenc}
\usepackage[utf8]{inputenc}
\usepackage[american]{babel}
\usepackage{epsfig}
\usepackage{graphicx}

\usepackage{hyperref}

\usepackage{booktabs}
\usepackage{multirow}
\usepackage{dcolumn}
\usepackage{amsmath}
\usepackage{mathtools}
\usepackage{amsfonts}
\usepackage{amssymb}
\usepackage{epstopdf}
\usepackage{bm}
\usepackage{siunitx}
\usepackage{braket}
\usepackage{enumitem}
\usepackage{soul}
\usepackage[table]{xcolor}
\usepackage{color}
\usepackage{transparent}

\usepackage{enumitem}
\usepackage{graphicx}
\usepackage[font=small]{caption}
\usepackage{subcaption}
\usepackage{pifont}

\begin{document}

\title{Cosmological tensions in the birthplace of the heliocentric model}

\author{Eleonora Di Valentino}
\email{e.divalentino@sheffield.ac.uk}
\affiliation{School of Mathematics and Statistics, University of Sheffield, Hounsfield Road, Sheffield S3 7RH, United Kingdom}

\author{Emmanuel Saridakis}
\email{msaridak@noa.gr}
\affiliation{National Observatory of Athens, Lofos Nymfon, 11852 Athens,
Greece}

\author{Adam Riess}
\email{ariess@stsci.edu}
\affiliation{Department of Physics and Astronomy, Johns Hopkins University, Baltimore, MD 21218, USA}
\affiliation{Space Telescope Science Institute, 3700 San Martin Drive, Baltimore, MD 21218, USA}

\date{\today}

\begin{abstract}
The theme of tensions in cosmology has become increasingly important in the cosmological community, proving capable of attracting new generations of scientists who want to be there and contribute to the next paradigm shift.
\end{abstract}

\maketitle

\section{Introduction}

With unexpectedly high participation both in-person and online, the “Tensions in Cosmology” Conference held by the Corfu Summer Institute and National Observatory of Athens in Greece, packed with talks over 4 days in September 2022, was a success. Covered by several international journalists, local TV, and a documentarian, and with the participation of more than 100 high profile speakers, including a Nobel Laureate, a Gold Medal recipient of the RAS, Gruber Cosmology Prize winners, etc, the conference explored both the well known Hubble constant $H_0$ and $S_8$ disagreements, now at more than 5$\sigma$ and 3$\sigma$ respectively, as well as a few more debatable tensions, such as large scale anomalies and curvature.  One speaker noted with amusement the similarity between discomfort with present tensions and past ``anti-Aristarchian'' efforts to disavow the Heliocentric model (in deference to the conference setting in Greece).

Evidence of tensions between the values of a few key cosmological parameters inferred in the late and early universe has been growing for a decade and has led many to wonder whether the simplest phenomenological model we have used for the last $\sim$ 25 years to describe our universe, i.e. the Lambda Cold Dark Matter model (LCDM), is complete.  Based on the recent literature and the conference attendance it is indisputable that the tensions in cosmology have become a topic of great interest for a growing number of early career researchers and PhD students, despite the skepticism of the older generation of (or more conservative) scientists. This growing segment of scientists do not believe the persistent evidence of cosmological tensions can reasonably be dismissed as unknown systematic errors in the data.   Most speakers thought our standard LCDM scenario of cosmology, a fitting model made of unknown components, exhibits cracks that are warranting serious study because they may point to the discovery of new physics or even a paradigm shift. Building on the success of the SNOWMASS Cosmology Intertwined White Paper \cite{Abdalla:2022yfr} co-authored by many of the participants, a takeaway conclusion of the conference was that we are now in the era of the cosmological tensions, and that if the 20th was the century of the particle physics, the 21st is the century of cosmology! 

\section{The cosmological tensions}

The conference opened with the presentation of the LCDM model and how it prospered for many years through an avalanche of data and the freedom of six parameters, until the divergent conclusions about $H_0$ and $S_8$ accumulated.  The most recent takes on $H_0$ are $H_0\sim 67.4 \pm 0.5$~km/s/Mpc, from the Planck collaboration using the Cosmic Microwave Background (CMB) in the early Universe and adopting a LCDM scenario, and $H_0\sim 73 \pm 1$~km/s/Mpc, from the SH0ES team calibrating Type Ia Supernovae (SNIa) with Cepheids. Improved calibrations of Cepheids from {\it Gaia} parallaxes of clusters and metallicity dependence were presented from two groups, which pushed this tension to $5.3\sigma$.  Talks were presented that corroborated this conclusion using alternative and complementary probes, such as Baryon Acoustic Oscillations (BAO) and ground-based CMB telescopes, or Surface Brightness Fluctuations (SBF), Masers and HII galaxies.  With so many independent probes making systematic-based explanations decreasingly plausible, the race has begun to find a theoretical solution that can revise or replace LCDM as well as alternative model-independent probes that can help to disentangle the problem.

Given the importance and significance of the Hubble tension, a number of presentations highlighted details of these measurements and tests for systematic errors. The Pantheon+ and SH0ES collaborations described their modeling of statistical and systematic uncertainties including a detailed covariance matrix with, for the first time, off-diagonal characterization. The Hubble constant constraints obtained with this 3-rung distance ladder method appeared robust with $\sim 70$ tests unable to lower $H_0$ below $\sim 72.5$~km/s/Mpc without throwing out data or producing new tensions. This measurement, now with a $1$~km/s/Mpc uncertainty, has proven resistant to attack, reanalysis and systematic effects due to its differential measurement, independent anchors, and sample matching. Also impressive were a 2-rung approach without SNIa, a 4-rung distance ladder including SBF, or a 3-rung approach replacing SNIa with HII galaxies, all in agreement with SH0ES. A first serendipitous look at Cepheids in a SNIa host presented from new James Webb Space Telescope (JWST) showed excellent agreement with the results obtained by Hubble Space Telescope (HST). Most attendees concluded it was no longer reasonable to defer this tension to vague speculations of possible ``unknown unknowns''. A number of results using  the Tip of the Red Giant Branch (TRGB) instead of Cepheids were presented with a range of $H_0$ from a low of $\sim70\pm2$~km/s/Mpc (CCHP collaboration) to a high of $\sim 77\pm6$~km/s/Mpc (ZTF) and a middle value of $73\pm3$~km/s/Mpc from TRGB+SBF.  Improved consistency between Cepheid and TRGB distances for the same hosts were presented and some remaining differences in $H_0$ from these studies were sourced to their treatment of SN Ia and the inclusion or absence of peculiar velocity and survey corrections. The Maser Cosmology Project presented $H_0=74 \pm 3$~km/s/Mpc - notably independent of Cepheids, TRGB, SBF and SNIa - and exciting plans for improvements from the resolution of the EHT. Strong lensing from quasar time delays confirmed the $H_0$ tension assuming conventional lens mass distributions, though became agnostic without these assumptions. 

Great hopes were placed in upcoming JWST programs using independent calibrations of SN Ia from Cepheids, TRGB, Miras and carbon stars. Emerging probes, such as cosmic chronometers, and strongly lensed quasars were discussed with more comprehensive quantifications of their systematic uncertainties, as well as the possibility of using quasars as standard candles.

Some discussions focused on the important role of the ground-based Atacama Cosmology Telescope (ACT) in the tensions panorama. A breakthrough in discrimination power is expected with the next ACT-DR6 release, a year away, that will improve the parameters constrained by the damping tail up to a factor $5$, and will scrutinize a recent hint of Early Dark Energy (EDE) in the ACT data.  This will be overtaken eventually by future experiments like the Simons Observatory and CMB-S4.
BAO and Redshift Space Distortion (RSD) data will continue to play an important role in the cosmological tensions debate, and the development of new approaches using large scale structure (LSS) to determine $H_0$ without the sound horizon will be useful to analyze the new DESI data.
The promise of Gravitational Wave Standard Siren (GWSS) cosmology, with and without the electromagnetic counterpart, also remains on the horizon, offering new methods and approaches.

The conference also discussed the other fascinating tension in cosmology today, the more than $3\sigma$ $S_8$ tension from the KiDS-1000 survey, or $> 4\sigma$ including other experiments. This appears to mimic the $H_0$ tension, whereby weak lensing, galaxy cluster and cluster count measurements in the local universe, seem to prefer a systematically different (lower) value than CMB+LCDM. 
Key points were the presentations of the Dark Energy Survey (DES) and KiDS-1000 results and their robustness, as well as their dependence on the different modelling choices and calibrations; great expectations are placed in their upcoming legacy releases. Also noteworthy was one past, prominent tension skeptic presenting a new take on the $S_8$ tension (``How I learned to stop worrying and love Cosmic Shear''), discussing how baryonic feedback or new dark matter physics on non-linear scales could be the source of this tension. And given that a bright future is expected with the Euclid Space Telescope, the Nancy Grace Roman Space Telescope and the Vera Rubin Observatory, the SNIa observed by these facilities can help study the $S_8$ tension. 

However, it appeared that the theorists were having the most fun at the meeting contemplating a Universe with tensions! We took part in an extensive discussion about the introduction of new physics, such as the indication at more than $3\sigma$ from ACT for an Universe of $\sim 10\%$ of EDE, or New EDE, that can alleviate the $H_0$ tension, though it does not improve the sibling $S_8$ tension nor is it favoured by the full Planck data, raising the stakes for ACT-DR6. An interesting possibility is also the variation of the fundamental constants, and in particular a varying electron mass, which was a favoured solution for the ``$H_0$ Olympics'' paper. Another front runner, interacting neutrinos, was discussed.  And as expected, we also reviewed arguments for and against modifications of General Relativity, or Modified Gravities of various classes (an evergreen proposal), since changing the engine of the Universe evolution is always a promising way for the explanation of new data, as demonstrated by the resolution of the famous tension between Mercury's perihelion data and the predictions of Newtonian gravity by another modified gravity theory, namely General Relativity.
More complicated late time solutions were proposed, such as dynamical and exotic Dark Energy possibilities, coupled and interacting Dark Matter and Dark Energy models, and running vacuum scenarios. Late time transitions models, as well as multi-parameters possibilities, in the inflationary scenario or in the curvature of the universe, were also presented. There were also those who have gone so far as to completely dismantle the LCDM standard model, testing its foundation and starting from scratch. However, at the moment none of the new theoretical solutions is compelling enough to clear the field, so the search for a new cosmological model, if needed, remains a work in progress.
Last but not least, was the proposal for new emerging data analysis techniques, such as the Artificial Neural Networks and Machine Learning, or the release of a new grid of parameters to perform model comparison and tension quantification.

\begin{figure*}
\centering
\includegraphics[width=0.8\textwidth]{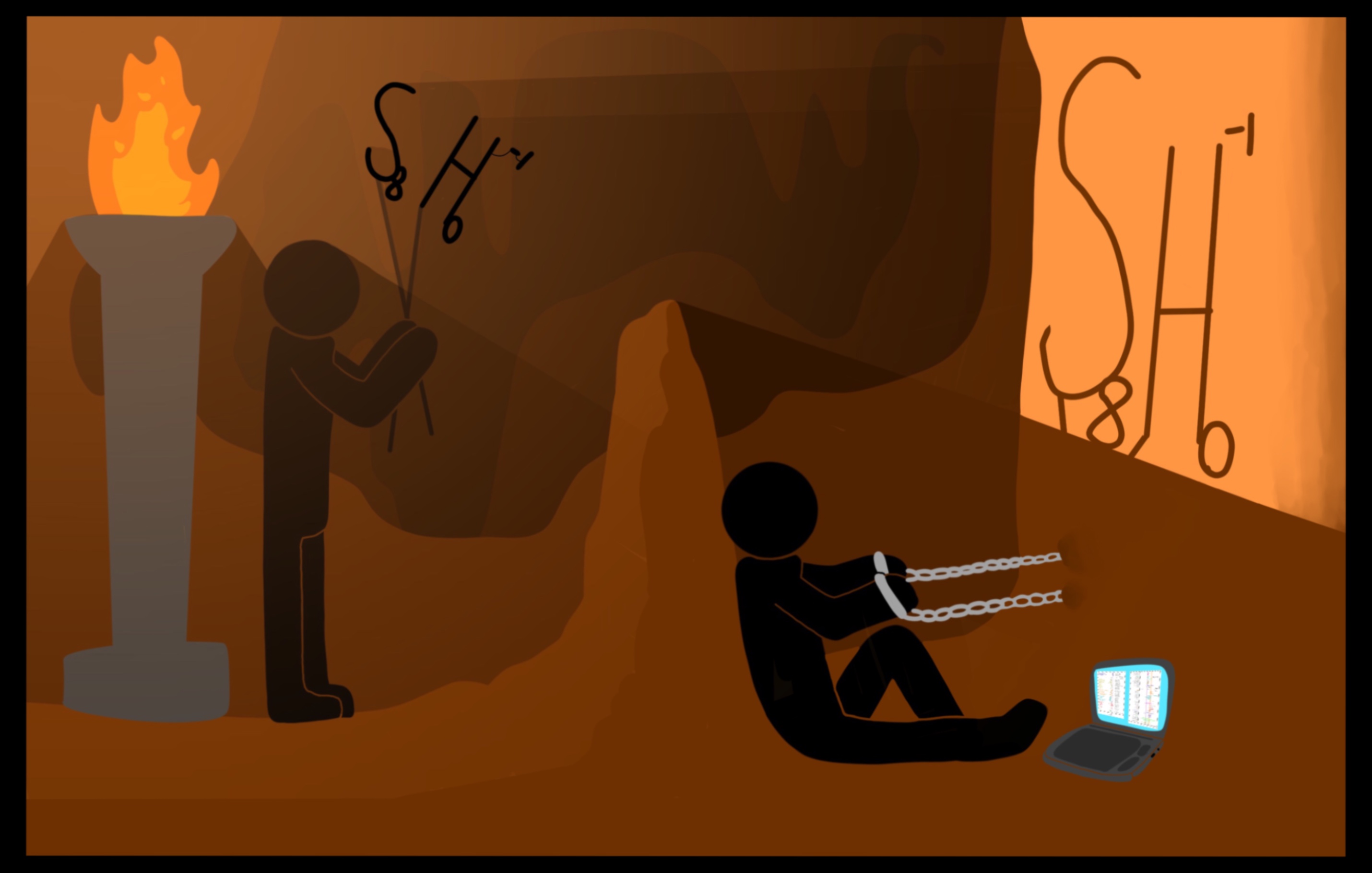}
\caption{As in the Allegory of the Cave, the observer of H0 and S8 is bound to one point of observation, so is unable to distinguish the shadows from the truth. Image credit: Tommaso Caprioli}
\label{fig}
\end{figure*}

Ultimately, what was the conclusion of this meeting? We have several powerful observations to compare, each relying on the current cosmological model, and they disagree. These observations have been deeply scrutinized for closed to a decade and we have further plans to improve these measurements in the coming years. But what if the tensions remain? Will we face the same outcome as in previous eras, where tensions between theoretical predictions and new data lead to new physics and to scientific paradigm shifts?
As presented by Plato in the Allegory of the Cave, we must be careful not to confuse the shadows we observe with the true reality. And the observations for the moment only show us the shadows, as graphically displayed in the Figure~\ref{fig}.  Hence, if too many points no longer match, given our assumptions, we will need to be ready to change those assumptions. And the news that emerged from the "Tensions in Cosmology" Conference is that the cosmological community has the necessary creativity to rise to the challenge.

\begin{acknowledgments}

EDV is supported by a Royal Society Dorothy Hodgkin Research Fellowship.
\end{acknowledgments}

\vfill
\end{document}